\documentclass[11pt]{article}
\usepackage[ hmargin=3 cm,vmargin=3cm]{geometry}

 \usepackage{graphicx}
 \usepackage{amsmath}
 \usepackage{amssymb}
  \usepackage{enumerate}
\usepackage{cite}

\usepackage{float}
\usepackage{tikz}
\usetikzlibrary{decorations.pathmorphing}
\usepackage[symbol]{footmisc}

\newcommand{\R}{\mathbb{R}}

\newcommand{\bey}{\begin{eqnarray}}
\newcommand{\eey}{\end{eqnarray}}
 \begin{document} \title{Relativistic time-of-arrival measurements: predictions, post-selection and causality problems}
 \author {Charis Anastopoulos\footnote{anastop@physics.upatras.gr}  \\
 {\small Department of Physics, University of Patras, 26500 Greece} \\
  and   \\
  Maria-Electra Plakitsi \footnote{plma101803@moraitis.edu.gr}\\
 {\small The Moraitis School, Psichiko, 15452, Greece }}
\maketitle
\begin{abstract}
We analyze time-of-arrival probability distributions for relativistic particles in the context of quantum field theory (QFT). We show that QFT leads to a unique prediction, modulo post-selection that incorporates properties of the apparatus into the initial state. We also show that an experimental distinction of different probability assigments is possible especially in near- field measurements. We also analyze causality in relativistic measurements. We consider a quantum state obtained by a spacetime-localized operation on the vacuum, and we show that detection probabilities are typically characterized by small transient non-causal terms. We explain that these terms originate from Feynman-propagation of the initial operation, because the Feynman propagator does not vanish outside the light-cone. We discuss possible ways to restore causality, and we argue that this may not be possible in measurement models that involve switching the field-apparatus coupling on and off.
\end{abstract}

\section{Introduction}

Finding a consistent quantum description for time-of-arrival measurements is a classic problem in the foundations of quantum theory. In the simplest time-of-arrival measurement, a particle is prepared on an initial state $|\psi_0 \rangle$ with positive momentum and  localized around $x = 0$. A particle detector is placed at $x = L$. The issue is to determine  the probability $P(t, L)dt$ that the detector clicks at  some moment between $t$ and $t+\delta t$. Despite the apparent simplicity of the problem, no unique time-of-arrival probability exists \cite{ML, ToAbooks}. Fundamentally, this is due to the fact that time is not a quantum observable. There is no self-adjoint operator for time \cite{Pauli}, hence, we cannot rely on Born's rule for a unique answer.

The time-of-arrival problem is exacerbated for relativistic particles, because it becomes entangled with another foundational problem, the issue of localization. Particle localization
 is essential to any description of measurements, because all particle-detection records  are localized in space and in time. However,  the existence of observables associated to spatial localization is in conflict with the requirement of causality, as evidenced by several theorems \cite{Schlieder, Malament, Heg1}.

The most well known set-up where this conflict appears is Fermi's two-atom problem. Fermi studied information transmission through a quantum field in a system of two localized atoms  at distance $r$ \cite{Fermi}. He  assumed that at time $t = 0$,  atom A is in an excited state and atom B is in the ground state. He asked when the influence of A will cause B to leave its ground state. In accordance with the locality principle that spacelike separated events cannot influence each other,
 he found that this happens only at time greater than $r$. However, Fermi's result was shown to be an artefact of an approximation  \cite{Shiro}. Later studies of the problem came to different results that were heavily dependent on approximations.  Eventually,  Hegerfeldt showed that the conflict of localization and causality is generic in relativistic systems \cite{Heg,Heg2}, it only requires  energy positivity and the treatment of atoms as localized in disjoint spatial regions------see, also the clarification of this result in \cite{Buch}.

In this paper, we analyze a broad class of probability distributions for the time of arrival of relativistic particles. This class was identified in Ref. \cite{AnSav19} through a analysis of measurements in QFT. Part of our motivation is the possibility of experimentally distinguishing   between different proposals for the time of arrival.
We also analyse the structure of small apparently super-luminal transient terms---analogous to the ones in the Fermi-atom problem---that appear in the time-of-arrival probability distribution.  We identify their origin and examine their implications for relativistic quantum theory of measurement.

An important reason for studying the time of arrival is that it forces us to re-conceptualize the description of quantum measurements. Ever since von Neumann \cite{vN}, measurements have been described as almost instantaneous processes that occur at a single moment of time $t$. In von Neumann measurements, the interaction of the apparatus  with the measured quantum system  is switched on for a pre-determined time interval---the exact timing of the measurement is determined by the shape of the switching function. Hence, the timing of the measurement is an external parameter  of the measurement scheme, and not a random variable of the experiment.
This logic has recently been extended to QFT measurements \cite{GGM22, FeVe}: the measured field and the apparatus are initially uncorrelated, and they interact only in a finite, predetermined spacetime region.

Time-of-arrival measurements challenge this measurement paradigm.  Actual particle detectors (e.g., photographic plates, silicon strips) have a fixed location in space and they are made sensitive for a long time interval during which particles may be detected. Therefore, the location of a detection event is a fixed parameter of the experiment;  the actual random variable is the detection time. Von Neumann type measurements are fundamentally incapable of describing time as a random variable, but it turns out that they can mimic some aspects of time-of-arrival measurements. However,  imitations have limitations: they work only to lowest order in perturbation theory and that it eventually leads to causality problems.

\medskip

Our results are the following.

First, we re-derive the relativistic time-of-arrival probabilities of Ref. \cite{AnSav19} in a von Neumann measurement scheme for quantum fields. The original derivation involved the Quantum Temporal Probabilities (QTP) method \cite{AnSav12, AnSav17, AnSav19, AnSav22, AnSavHu22} that explicitly constructed a probability density with respect to time. In the von Neumann measurement scheme, we use a switching function that is localized in a compact spacetime region, and we reinterpret the probabilities in order to define a probability density. This works only to leading order in perturbation theory. We use this alternative derivation in order to identify the problems that persist in this treatment of measurements.

Time-of-arrival probability distributions are post-selected, i.e., they refer only to the fraction of particles that has been detected. They depend on the apparatus through a specific operator $\hat{S}$ that describes the localization of the detection records. We show that this operator can be absorbed into a post-selection of the initial state.  {\em A unique time-of-arrival probability measure} ensues,  modulo post-selection. This measure was first derived by Leon \cite{Leon} and then rederived from a QFT analysis in \cite{AnSav12}. In the non-relativistic limit, it coincides with Kijowski's time-of-arrival distribution \cite{Kijowski}.

The measured time of arrival probability distribution does depend on the properties of the apparatus. We analyze the probability measure for an operationally meaningful class of initial states that was recently proposed, in relation to experiments for distinguishing between different time-of-arrival proposals.  We find that the different distributions are in principle distinguishable in near-field experiments.

Then, we analyse locality in the time-of-arrival probabilities. We consider an initial state that is generated by a localized external source acting on the quantum field vacuum. By "localized" we mean that the source has  support in a compact spacetime region. We find that the time-of-arrival probability has a small but non-zero contribution outside the light-cone of the source's support. We analyze the origin of this term, and we find that it originates from the fact that in QFT sources evolve with the Feynman propagator, and Feynman propagator is non-zero outside the light-cone. We argue that  von-Neumann type models, that rely on switching on the interaction, may not be able to resolve this problem.

Finally, we briefly revisit the QTP description of relativistic measurements, and present possible strategies through which the super-luminal transient terms can be consistently removed.

\section{ Relativistic time-of-arrival probabilities }
In this section, we first re-derive the time-of-arrival probabilities of Ref. \cite{AnSav19} using a von Neumann type of measurement. We show that these probabilities are unique modulo post-selection, and that the effect of the detector---as determined by its localization properties---can be distinguished in near-field measurements.

\subsection{Detection probability from a von Neumann type measurement}
We consider the measurement of a free scalar field $\hat{\phi}(x)$ on a Hilbert space ${\cal F}$, interacting with an apparatus described by a Hilbert space ${\cal H}$. As our focus is time-of-arrival measurements, we will work only in one spatial dimension. The reason is that only particles that propagate along the axis that connects the source with the detector contribute to the total probability.

 The Hamiltonian of the total system is
\bey
\hat{H} = \hat{H}_{\phi}\otimes \hat{I} + \hat{I} \otimes \hat{H}_A + \hat{H}_{I},
\eey
where $\hat{H}_{\phi}$ is the quantum field Hamiltonian and $\hat{H}_A$ the Hamiltonian of the apparatus.  We consider a von-Neumann type of measurement, in which the interaction is switched on for a finite time. Then, we choose an interaction Hamiltonian
\bey
\hat{H}_I = \int  dx F_{\bar{t}, \bar{x}}(t, x)\hat{\phi}(x) \otimes \hat{J}(x),
\eey
where $F_{\bar{t}, \bar{x}}(t, x)$ is a switching function centered around the spacetime point $(\bar{t}, \bar{x})$, and $\hat{J}(x)$ is a current operator on ${\cal F}$\footnote{Certainly, the idea of a switching interaction is not realistic in QFT. Fields interact with the apparatuses through terms defined by the Standard Model of particle physics, and in a Poincar\'e covariant theory, the interaction terms are always present.}.

We assume an initial state $|\psi\rangle$ for the field and an initial state $|\Omega\rangle$ for the detector. It is convenient to identify $|\Omega\rangle$ with the ground state of the Hamiltonian $\hat{H}_A$, $\hat{H}_A |\Omega \rangle = 0$, and also to be annihilated by the generator of space translations $\hat{P}_A$ of the apparatus. Then, the probability that the detector is found in an excited state, when measured after the interaction has been switched off,  is given by
\bey
\mbox{Prob}(\bar{t}, \bar{x}) = \langle \psi_0 ,  \Omega|  \hat{S}^{\dagger}_{\bar{t}, \bar{x}} (\hat{I}\otimes \hat{I} - |\Omega\rangle \langle \Omega|)   \hat{S}_{\bar{t}, \bar{x}}  |\psi_0, \Omega\rangle, \label{probtx}
\eey
expressed in terms of the S-matrix $\hat{S}_{\bar{t}, \bar{x}} = {\cal T} \exp\left[- i \int dt dx  F_{\bar{t}, \bar{x}}(t, x) \hat{\phi}(t,x) \otimes \hat{J}(t,x) \right]$; $\hat{\phi}(t, x)$ and $\hat{J}(t, x)$ are Heisenberg-picture operators and ${\cal T}$ stands for time-ordering.

To leading order in perturbation theory,
\bey
\mbox{Prob}(\bar{t}, \bar{x}) = \int dt_1 dx_1   dt_2 dx_2  F_{\bar{t}, \bar{x}}(t_1, x_1)  F_{\bar{t}, \bar{x}}(t_2, x_2) G(t_x,t_1; t_2,x_2) \langle \Omega|\hat{J}(t_1, x_1) \hat{J}(t_2, x_2)|\Omega\rangle, \label{probX}
\eey
where
\bey
G(t_1,x_1; t_2,x_2)= \langle \psi_0|\hat{\phi}(t_1,x_1) \hat{\phi}(t_2, x_2)|\psi_0\rangle \label{propagator}
\eey
is a two-point correlation function for the field.

We take  a reference point $x_0 = 0$ on the apparatus, and we define $|\omega\rangle = \hat{J}(0) |\Omega\rangle$. Then,
 we can write  $\Omega|\hat{J}(t_1, x_1) \hat{J}(t_2, x_2)|\Omega\rangle = R(t_2 - t_2, x_2 - x_1)$, where
 \bey
 R(t, x) = \langle \omega|e^{i\hat{H}_At - i \hat{P}_Ax}|\omega\rangle \nonumber
 \eey
is the detector kernel. The key property of $R(t, x)$ is that by energy positivity, its Fourier transform
\bey
\tilde{R}(E, K) := \int dt dx R(t, x) e^{-iEt +iKx} = 2 \pi \langle \omega|\delta(E-\hat{H}_A) \delta(K - \hat{P}_A)
\eey
vanishes for  $E < 0$.

In Eq. (\ref{probX}) $\bar{x}$ and $\bar{t}$ appear as parameters, not as random variables. However, Eq. (\ref{probX}) has a natural interpretation in terms of a  probability density. Consider a homogeneous
switching function
$F_{\bar{t}, \bar{x}}(t, x) = f( t - \bar{t}, x - \bar{x})$, where $f(t, x) = \exp\left[- \frac{t^2}{2\delta_t^2} - \frac{x^2}{2 \delta_x^2}  \right]$ for space and time spreads $\delta_x$ and $\delta_t$, respectively. Gaussians satisfy the identity
 \bey
f( t, x) f(t', x') = f^2\left(\frac{t+t'}{2}, \frac{x+x'}{2}\right) \sqrt{f}(t - t', x-x'). \label{gauidty}
\eey
Then, Eq. (\ref{probX}) takes the form
\bey
\mbox{Prob}(\bar{t}, \bar{x}) = \int dt dx f^2\left(\bar{t} - t, \bar{x}-x\right) P(t, x),
\eey
i.e., it is a convolution of the probability density
\bey
P(t, x) = \int ds dy R(s, y) \sqrt{f}(s, y)G(t- \frac{s}{2}, x - \frac{1}{2}y; t +\frac{1}{2}s, x +\frac{1}{2}y). \label{ptx2}
\eey
Typically, the detector kernel decays to zero for $|t|$ larger than a temporal scale $\hat{\sigma}_T$ or $|x|$ larger than a length scale $\sigma_x$. Both scales depend on properties of the apparatus. Taking $\delta_t >> \sigma_t$ and $\delta_x >> \sigma_x$, the contribution of $\sqrt{f}$ can be dropped from Eq. (\ref{ptx2}), so that
\bey
P(t, x) = \int ds dy R(s, y) G(t- \frac{s}{2}, x - \frac{1}{2}y; t +\frac{1}{2}s, x +\frac{1}{2}y). \label{ptx3}
\eey
It is important to emphasize that the probability density (\ref{ptx3}) is meaningful only to leading order in perturbation theory. We cannot construct a spacetime density out of higher orders terms in this procedure.  This is to be constructed with the QTP method, briefly described in Sec. 3, that leads to probability densities at all orders of perturbation theory.

 Eq. (\ref{ptx3}) holds for any scalar field theory. For a free field of mass $m$,
\bey
\hat{\phi}(t, x) = \int \frac{dp}{2\pi \sqrt{2\epsilon_p}} [\hat{a}(k)e^{ipx - i \epsilon_pt} + \hat{a}^{\dagger}(k)e^{-ipx + i \epsilon-pt}],
\eey
where $\epsilon_p = \sqrt{p^2+m^2}$.

 Eq. (\ref{ptx3}) becomes
\begin{eqnarray}
P(t, x) = P_0 + \int \frac{dpdp'}{2\pi } \frac{\rho(p,p')}{2\sqrt{\epsilon_p \epsilon_{p'}}} \; \tilde{R}\left( \frac{p+p'}{2}, \frac{\epsilon_p + \epsilon_{p'}}{2}\right) e^{i(p-p')x - i (\epsilon_p - \epsilon_{p'})t}, \label{ptx4}
\end{eqnarray}
where  $\hat{\rho}(p,p') =  \langle \psi|\hat{a}^{\dagger}(p')\hat{a}(p)|\psi\rangle$ is the single -particle reduced density matrix. The term
\bey
P_0 = \int \frac{dp}{4\pi \epsilon_p} \tilde{R}(p, \epsilon_p) \label{p0}
\eey
 is constant and state-independent. It corresponds to vacuum noise, i.e., a background rate of false alarms. Hence, a detection signal exists only as long as $P(t, x)$ is larger than $P_0$. We will ignore $P_0$ in what follows, except for checking whether  transient terms that appear in $P(t, x)$ are actual detection signals.

\subsection{Post-selection with respect to recorded events}

Eq. (\ref{ptx4}) is an unnormalized probability density. To normalize, we consider a set-up where the particle source is in the vicinity of $x = 0$, and the detector at $x = L$, where $L > 0$ is a macroscopic distance. In this setup, the contribution of negative momenta to $P(t, L)$ is negligible. It is therefore convenient to normalize over initial states with support only over positive momenta. Then, integrating over $t \in \R$, we obtain the total detection probability for constant $L$,
\bey
P_{tot} = \int dp \rho(p, p)\; \frac{\tilde{R}(p, \epsilon_p)}{2 p}. \label{posts}
\eey
 This means that
\bey
\alpha(p) = \frac{\tilde{R}(p, \epsilon_p)}{2 p}
\eey
is the absorption coefficient of the detector, i.e., it records the fraction of particles of momentum $p$ that are absorbed, and hence detected. Note that we integrated $t$ over the full real axis, because the contribution of $t < 0$ is very small if the initial state has no negative momenta. This is an approximation, as only positive values of $t$ are physically meaningful.

The conditional probability distribution $P_c(t, L):= P(t, L)/P_{tot}$ is normalized to unity, as long as the initial state contains positive momenta. Then, we define the post-selected density matrix
\bey
\rho_{ps}(p, p') := \frac{\sqrt{\alpha(p)} \sqrt{\alpha(p')}}{P_{tot}} \rho(p,p'). \label{rpp1}
\eey
Note, that for an initial pure state, the post-selected state remains pure.
Then,
\begin{eqnarray}
P_c(t, L) = \int \frac{dpdp'}{2\pi } \rho_{ps}(p,p')  \sqrt{v_p v_{p'}} S(p,p') e^{i(p-p')L - i (\epsilon_p - \epsilon_{p'})t}, \label{ptxb}
\end{eqnarray}
where $v_p = p/\epsilon_p$ is the particle velocity.  $S(p, p')$ are the matrix elements $\langle p|\hat{S}|p'\rangle$ of the {\em  localization operator} $\hat{S}$, defined by
\begin{eqnarray}
 \langle p|\hat{S}|p'\rangle  := \frac{\tilde{R}\left( \frac{p+p'}{2}, \frac{\epsilon_p + \epsilon_{p'}}{2}\right)}{\sqrt{\tilde{R}(p, \epsilon_p) \tilde{R}(p', \epsilon_{p'})}}.  \label{lpp}
\end{eqnarray}
By definition, $\langle p|\hat{S}|\hat{p'}\rangle \geq 0 $ and  $S(p, p) = 1$.  Its name originates from the fact that $\hat{S}$ describes the localization of an elementary measurement event. $\hat{S}$ is a positive operator if $\ln \tilde{R}(p, \epsilon_p)$ is  a  convex  function of $p$. Then, the Cauchy-Schwarz inequality applies,
\begin{eqnarray}
\langle p|\hat{S}|p'\rangle \leq \sqrt{\langle p|\hat{S}|p\rangle \langle p'|\hat{S}|p'\rangle} =  1. \label{CSa}
\end{eqnarray}
Maximal localization is achieved when  Eq. (\ref{CSa}) is saturated, i.e.,  for  $\langle p|\hat{S}|p'\rangle =  1$. For a pure initial state
\bey
P_c(t, L) = \left| \int_0^{\infty} dp \psi_{ps}(p) \sqrt{v_p} e^{ipL - i \epsilon_pt}\right|^2. \label{pcx}
\eey
Eq. (\ref{pcx}) was first derived by \cite{Leon}. In the non-relativistic limit, it coincides with Kijowski's formula \cite{Kijowski}. Note that these expressions differ from those obtained by the current operator $\hat{J} = \frac{1}{2} \left[\hat{v} \delta(\hat{x} - L) +  \delta(\hat{x} - L) \hat{v} \right]$,
\bey
P_J(t, L) =  \int \frac{dpdp'}{2\pi } \rho_{ps}(p,p')  \frac{1}{2}(v_p + v_{p'})  e^{i(p-p')L - i (\epsilon_p - \epsilon_{p'})t}. \label{probcurr}
\eey
Note that Eq. (\ref{probcurr}) does not, in general guarantee, the positivity of probabilities.

To understand the meaning of the localization operator, we evaluate its
 Wigner-Weyl transform.
\bey
\tilde{S}(x, p) :=   \int \frac{d\xi}{2\pi} S\left(p-\frac{\xi}{2}, p + \frac{\xi}{2}\right) e^{-i\xi x},
\eey
It is straightforward to show that $\int dx \tilde{S}(x,p) = S(p, p) = 1$. Furthermore,
 by Bochner's theorem, $\tilde{S}(x, p) \geq 0$. Hence, $\tilde{S}(x, p)$ is a $p$-dependent probability distribution with respect to $x$, which we will denote by $u_p(x)$. This probability distribution defines the irreducible spread in determining the location of the measurement record. Maximum localization corresponds to $u_p(x)  = \delta (x)$ \cite{AnSav19}. We note that the general form (\ref{ptxb}) is fixed by the requirements of Poincar\'e covariance \cite{Wer86}; however, this requirement does not fix the properties of the localization operator or its physical interpretation in terms of properties of the apparatus.

We can absorb the localization operator $\hat{S}$ through further post-selection of the initial state,
\bey
\tilde{\rho}_{ps}(p, p') = S(p, p') \rho_{ps}(p,p'). \label{reff}
\eey
It is straightforward to show that $\tilde{\rho}_{ps}$ is a density matrix. Hence, all time-of-arrival probabilities can be brought in the form (\ref{CSa}) modulo post-selection of the initial state.

However, this uniqueness result is primarily structural, and it is not reflected in the measured probability distributions, which depend strongly on the properties of the detector. The latter are incorporated   in two quantities, the absorption coefficient $\alpha(p)$ and the localization operator $\hat{S}$. In principle, we can determine both for a given detector: $\alpha(p)$ is essentially determined by the attenuation coefficient of the detecting medium, while $\hat{S}$ can be determined through time-of-arrival experiments. Hence, both redefinitions (\ref{reff}) and (\ref{rpp1}) make operational sense.

\subsection{Wigner representation}
We gain some insight into the structure of the time of arrival probabilities by expressing them in terms of the Wigner function associated to the density matrix $\rho_{ps}(p,p')$,
\bey
W(x,p) = \int \frac{d\xi}{2\pi} \rho_{ps}\left(p-\frac{\xi}{2}, p + \frac{\xi}{2}\right) e^{-i\xi x}.
\eey
With these definitions,  Eq. (\ref{ptxb}) becomes
\begin{eqnarray}
P_c(t, L) = \int dp dx_0 dx_f W(x_0, p) u_p(x_f) F_t(L - x_f - x_0, t), \label{pct3}
\end{eqnarray}
where
\bey
F_t(x,  p) =\int_{-2p}^{2p} d \xi \sqrt{|v_{p+\xi/2} v_{p- \xi/2}|} e^{i x \xi - i(\epsilon_{p +\xi/2} - \epsilon_{p - \xi/2})t} \label{ftxp}
\eey
is a function on the classical phase space that represents the quantum time-of-arrival observable.

Changing variables to $x := x_f - x_0$, we can write Eq. (\ref{pct3}) as
\bey
P_c(t, L) = \int dp dx \tilde{W}(x, p) F_t(x,p),
\eey
where $\tilde{W}(x, p) = \int dx_f u_p(x_f) W(x - x_f, p)$ is the  Wigner function associated to the post-selected quantum state $\tilde{\rho}_{ps}$. Hence, the localization operator implements a position smearing of the particle's Wigner function.

In the non-relativistic limit,
\bey
F_t(x,p) = m^{-1} \int_{-2p}^{2p} d \xi \sqrt{p^2 -\frac{\xi^2}{4}} e^{ix \xi - i p \xi t/m} = v_p \frac{ \pi J_1(2p|x - v_pt|)}{|x - v_pt|}, \label{nonr1}
\eey
where $J_1$ is the Bessel function. Note that $F_t$ takes negative values. This  is fully compatible with the positivity of the time-of-arrival probabilities, because probability distributions concentrated in the regions where $F_t$ is negative are not admissible Wigner functions for quantum states.

\begin{figure}
 \includegraphics[height=6cm]{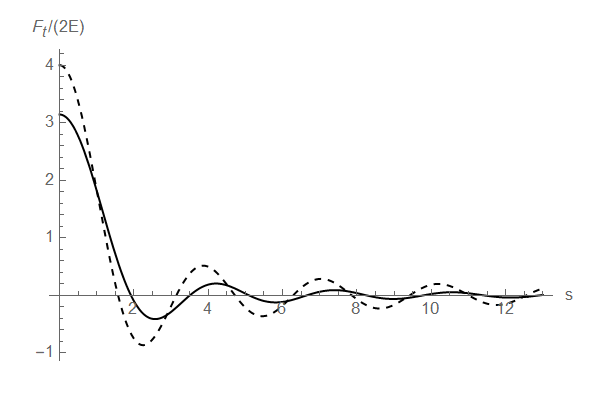}
    \caption{ The time-of-arrival functions (\ref{nonr1}) (solid) and (\ref{nonr2}) (dashed)  divided by $2E = p^2/m$, as  a function of $s:= p(x - v_pt)$.  }
    \label{fplot}
\end{figure}

 In the ultra-relativistic limit ($m = 0$),
\bey
F_t(x,p) = \int_{-2p}^{2p} d \xi e^{i x \xi - i \xi t} = \frac{\sin[2p(t-x)]}{p(t-x)}
\eey
These expressions are to be compared with the corresponding classical expression for the time-of-arrival observable,
\bey
F^{cl}_t(x, p) =  \delta( t - x/v_p),
\eey
and the expression obtained from the non-relativistic current operator
\bey
F^{J}_t(x, p) = \int_{-2p}^{2p} d \xi \frac{1}{2} (v_{p+\xi/2} + v_{p- \xi/2}) e^{i x \xi - i(\epsilon_{p +\xi/2} - \epsilon_{p - \xi/2})t} =  v_p  \frac{2 \sin[2p(x- v_p t)]}{(x - v_pt)} \label{nonr2}
\eey
The different versions of the time-of-arrival functions in the non-relativistic regime are plotted in Fig. \ref{fplot} as a function of $s:= p(x - v_pt)$. The classical time of arrival observable is a delta function at $s = 0$. Eqs. (\ref{nonr1}) and (\ref{nonr2}) have finite spread around $s = 0$, and they differ strongly near $s = 0$.

\subsection{Probability dependence  on the detector kernel}

One motivation of this paper is the possibility of experimental comparison between different proposals for the time-of-arrival probabilities. To do so, we need to specify an operationally meaningful initial condition. Such a condition was proposed in \cite{DaDu}, where it was suggested that a strong divergence of the prediction by Bohmian mechanics from the predictions of the semi-classical theory for the time of arrival.

\begin{figure}
 \includegraphics[height=8cm]{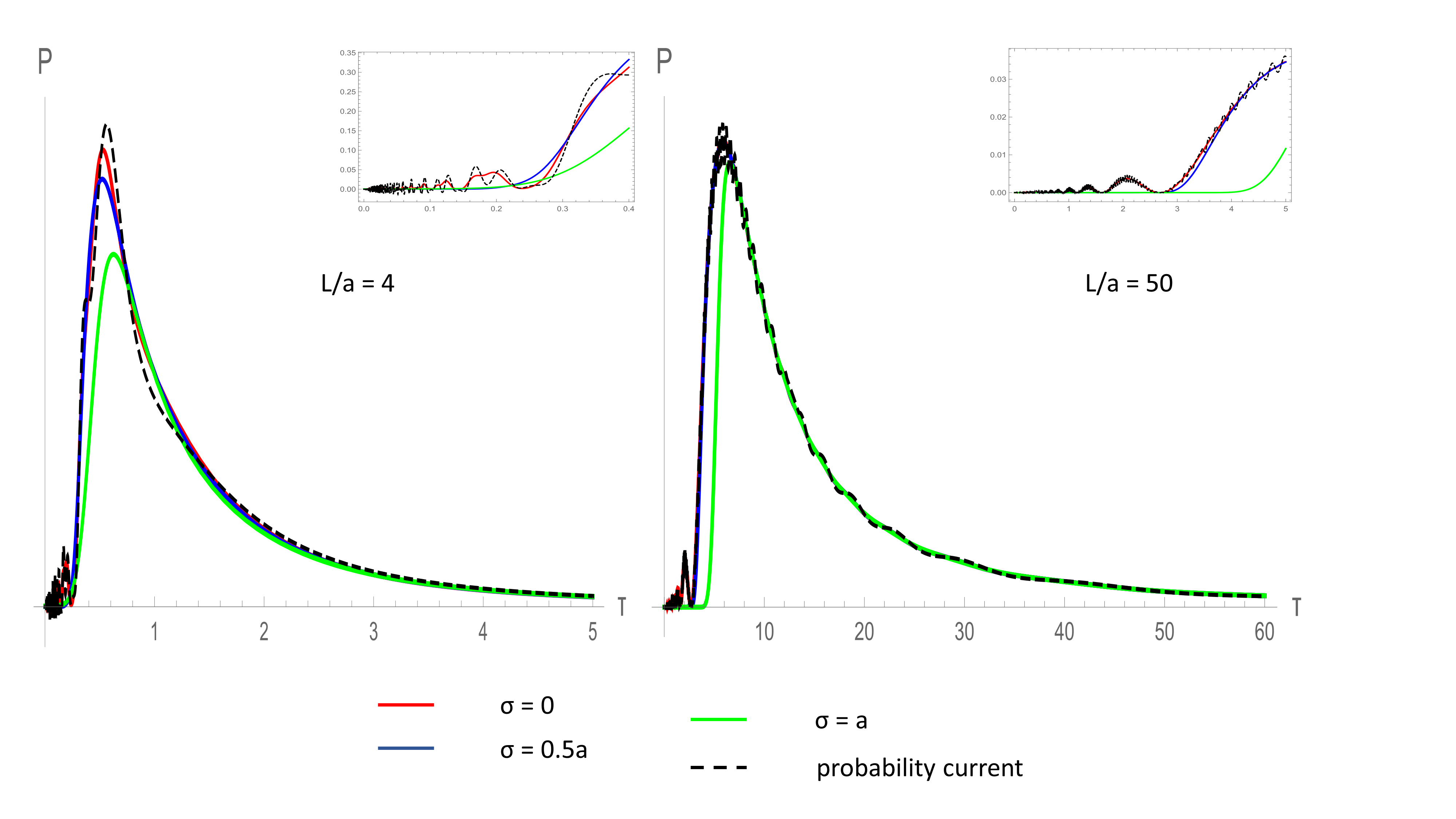}
    \caption{ The probability density $P_c(\tau)$ as a function of $\tau:= t/(2ma^2)$,  for different values of the localization length $\sigma$ and for different distances, and for the pseudo-probability density (\ref{probcurr}). The plots correspond to the non-relativistic regime. The insets show the early-time behavior of $P_c(t)$.}
    \label{pplot}
\end{figure}

The idea is to prepare a number of particles in a box of width $a$, and we remove one wall of the box (or both) at time $t = 0$. Then, the initial state is an eigenfunction of the Hamiltonian for a particle in a box,
$\psi_n(x) = \sqrt{\frac{2}{a}} \sin(\pi x/a_n)$ for $x \in [0, a]$ and zero otherwise; here $a_n = a/n$ and $n = 1, 2, \ldots$. Equivalently, in the momentum representation
\bey
\psi_n(p) = \pi \sqrt{\frac{2}{a}}  a_n \frac{1+e^{-ia_np}}{\pi^2 - a_n^2 p ^2}.
\eey
To understand the dependence of the probability density $P_c(t)$ on the detection kernel, we consider a simple model, where $S(p, p') = \exp[-\sigma^2 (p-p')^2]$, i.e., a detection kernel that is determined by the localization length $\sigma$. In Fig. \ref{pplot}, we plot $P_c(t)$ for different values of $\sigma$; we also plot the pseudo-probability density (\ref{probcurr}). We see that the  probability densities differ significantly at early times, and they have the same asymptotic fall-off behavior.

We also find that if $\sigma/\alpha << 1$, $P_c(t)$ essentially coincides with the maximal-localization form (\ref{pcx}). The distribution (\ref{probcurr}) becomes negative at early times, hence, it has no operational significance. We also note that the differences are more pronounced for short $L$, i.e., in the near-field regime. At large $L$, only distributions with $\sigma/\alpha$ of order one or larger are distinguishable.

\section{Causality issues}
In this section, we show that the time-of-arrival probabilities suffer from small transient super-luminal terms. These terms originate from the fact that unitary evolution implies Feynman-propagation of sources, and the Feynman propagator does not vanish outside the light-cone. We discuss possible strategies for removing such terms from the theoretical description, and we argue that this may not be possible in von Neumann-type models of measurement.

\subsection{Apparent causality violation}
Any theory of relativistic measurements for quantum systems must respect causality, in the sense that if the system is initially prepared within a localized spacetime region $A$, then there should be no detection signal in any region $B$ spacelike to $A$.

To implement this condition, we must first identify initial states that are localized in  a spacetime region. This is a difficult task. First, we know that spatial localization at a moment of time does not work. Any notion of spatial localization in the single-time quantum state (even approximate ones)  leads to faster-than-light signals  \cite{Malament, Heg1}. This conclusion is not an artefact of specific models, but a consequence of fundamental properties of relativistic quantum systems, namely, Poincar\'e covariance and energy positivity.

Rather than spatially localized states, we consider spacetime localized operations. We assume that the field is initially in the vacuum state $|0\rangle$, and that we prepare an initial state for the time-of-arrival measurement by an external intervention. This intervention has support in a compact spacetime region that lies wholly before the Cauchy surface
$t = 0$. Then, we take the resulting state as $|\psi_0\rangle$ in Eq. (\ref{propagator}).

\begin{figure}
 \includegraphics[height=8cm]{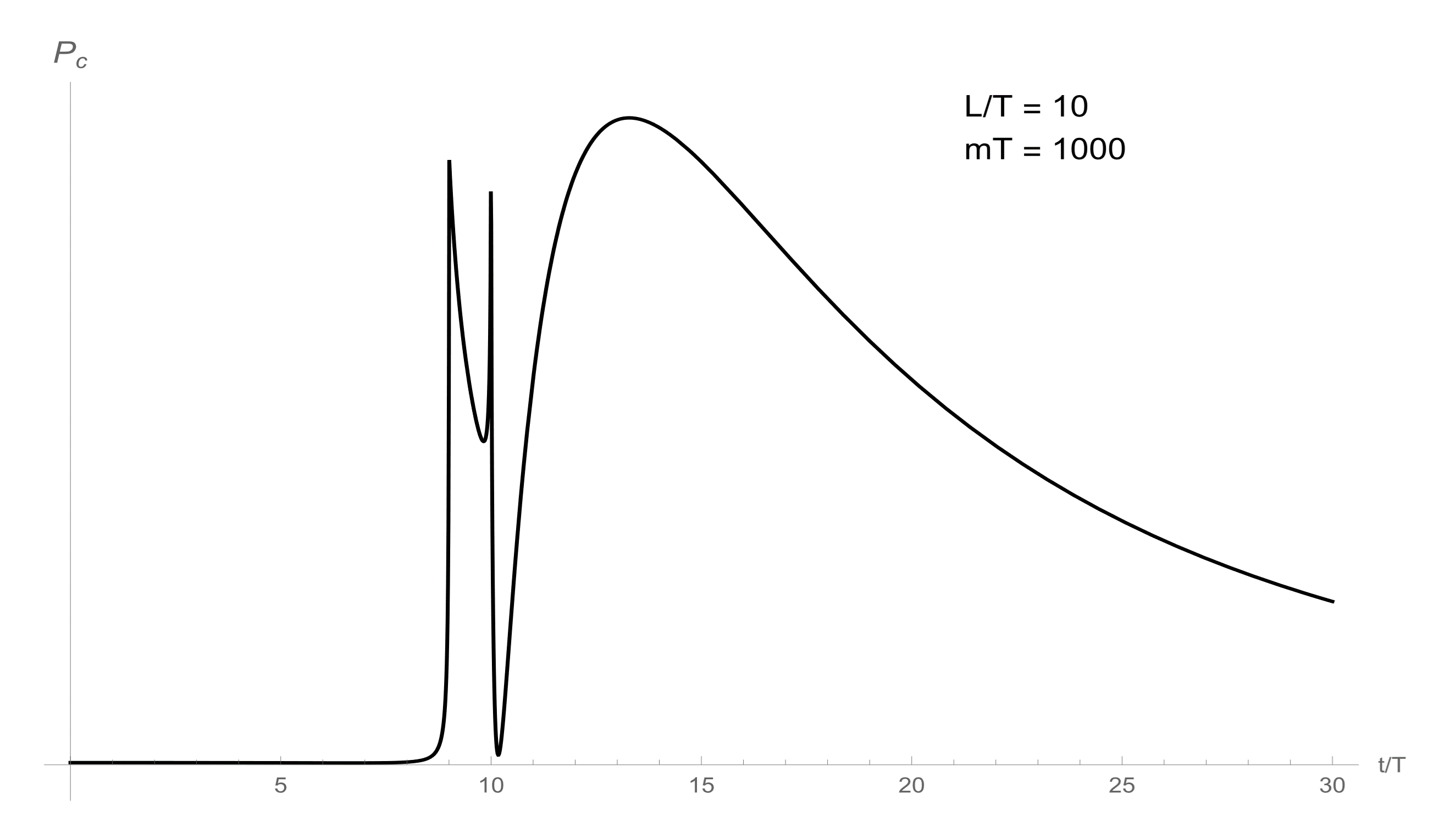}
    \caption{ The probability density $P_c(t)$ for maximal localization as a function of $\tau:= t/T$,  for $L/T = 10$ and $mT = 10^3$. The non-causal behavior is manifested in the jump of the probability density prior to $t = L - T$.}
    \label{plotpc}
\end{figure}

The simplest type of action on a field involves the switching on of a source, i.e., including a time dependent term $\int dx \hat{C}(x) J(x, t)$, where $\hat{C}(x)$ is a local composite operator and $J(t, x)$ a source with support in a compact spacetime region. Then, $|\psi_0\rangle = {\cal T} \exp[-i \int dt dx \hat{C}(x, t) J(x,t)] |0\rangle$. For $\hat{C}(t, x) = \hat{\phi}(t, x)$, $|\psi_0\rangle$ coincides with the field coherent state $e^{-i \int dt dx J(x, t) \hat{\phi}(t, x)}|0\rangle$, modulo a phase. Hence, the single-particle wave function is identified with
\bey
\psi_0(p) = \frac{1}{\sqrt{2\epsilon_p}} \int dt dx J(t, x)e^{-ipx + i \epsilon_pt}. \label{psi0j}
\eey
Note that states of this form cannot have support only on positive momenta, hence, the probability density $P_c(t)$ is not normalized to unity\footnote{Any state with only positive momenta, has a tail in position that extends up to the location of the detector, and is therefore not appropriate for testing causality.}.
We consider an source that vanishes outside the spatial interval $[-a, 0]$ and $[-T, 0]$, where $a >0$ and $T > 0$. We work in the regime $a << T$, i.e., a spatially localized interaction of long duration. For concreteness, we take
\bey
J(t, x) = \left\{ \begin{array}{cc} \delta(x) \sin (\pi t/T), &t \in [ -T, 0]\\0,& \mbox{otherwise}\end{array}\right..
\eey
 Then,
\bey
\psi_0(p) =- \frac{\pi T (1 +e^{-i\epsilon_p T})}{\pi^2 - \epsilon_p^2 T^2}.
\eey
 Causality requires that the detection probability at $x = L$ vanishes for all $ t < L - T$, so that the detector is only influenced by events that happen after we started our intervention.
 This is not the case. Fig. \ref{plotpc} shows that $P_c(t)$ starts becoming appreciably larger than zero slightly before $t = L - T$, and in fact, it is non-zero for all times $t > 0$. The non-causal component is more pronounced in the near field regime, i.e., small $L$.

This result is remarkable, but, given past results on apparent causality violation in localized relativistic systems,  not unexpected. The specific context of this apparent causality violation is rather different. Past results focus on the evolution of spatially localized quantum states at a single moment of time. Here, we considered a {\em time-extended} external localized intervention.   Here, the problem is not an artefact of using idealizations from non-relativistic theory in describing a fundamentally relativistic system. It is also not an artefact of perturbation theory. The conditional probability density $P_c(t)$ that manifests the causality violating terms is of order $\lambda^0$ with respect to the field-apparatus coupling $\lambda$, higher perturbative corrections are negligible for sufficiently weak coupling.

This result does not indicate a fundamental failure of causality, rather it points at the limitations of von Neumann's description of measurements, i.e., the description of measurements in terms of an interaction that is switched on. If we take the quantum formalism literally, the probability density (\ref{probtx}) does not describe a measurement that takes place at time $\bar{t}$; $\bar{t}$ specifies the moment when the field-apparatus is on. Rather, Eq. (\ref{probtx}) refers to a measurement that takes place after the interaction has been switched-off. This distinction may be unimportant when the interaction time is very small. In this formulation, there is no lower limit on the  width of the switching function, we can consider a delta function that is  proportional to
$\delta(t - \bar{t})$. If we take the duration of the interaction larger than the distance between source and detector, no causality problem arises. However, in a measurement theory, the switching function is supposed to model a defining feature of the apparatus, it cannot change arbitrarily if we decide to place the source at a different distance.

In S-matrix theory the coupling is assumed to be switched on and off adiabatically. This means that the effective duration of the interaction is arbitrarily large,  and measurement takes place effectively at $t \rightarrow \infty$. In this description, transient terms disappear.
However, this is not a solution of the problem, merely a sidestep. In any particle detector, particles are recorded at specific moments of time, i.e., particle detection is a temporally localized process. An adiabatic switching of the interaction fails to localize the measurement event in time. Hence,  unless we accept causality violations, we must admit that measurement models that rely on switching interactions  cannot account for a fundamental observed  feature of the particle detection process, namely, that it is localized in time.

One possible way of restoring causality is by showing that the causality violating terms are impossible to measure, as a matter of principle. This may be possible, because QFT measurements are invariably characterized by false alarms, i.e., spurious detection events. Such events essentially define an irreducible background noise that persists even if the field is in the vacuum state. A detection signal exists only if it leads to a significant spike of recorded events over the noise background. This means that the super-luminal transient terms may be ignored if they contribute to probability less than the noise term due to vacuum.

In the present model, this noise is expressed by the constant background $P_0$ of Eq. (\ref{p0}). This term is quite strong, in fact, it drowns a large part of the signal in the ultra-relativistic regime, and not only the transient terms. However, the strength of this term is an artefact of coupling switching. A realistic model for the apparatus must take into account that the initial state of the apparatus is correlated with the field vacuum, i.e., it is a dressed state. The noise $P_0$ incorporates also the equilibration of the apparatus with the field vacuum when the coupling is switched on. This equilibration is a spurious process. In the full description, the initial state of the apparatus is already equilibrated with the field. This implies that the restoration of causality through a signal/noise analysis requires a different approach to measurements. In Sec. 3.1., we argue that such an analysis is feasible in the QTP approach.

\subsection{Retarded propagator versus Feynman propagator}
For the initial state (\ref{psi0j}), the detection probability becomes
\bey
P(t, x) = P_0 + \int ds dy R(s, y) \left[ 2 \mbox{Re} \left[ (\Delta_FJ)(t- \frac{s}{2}, x - \frac{1}{2}y) (\Delta_FJ)( t +\frac{1}{2}s, x +\frac{1}{2}y)\right]
\right. \nonumber \\
\left. +
(\Delta_FJ)(t- \frac{s}{2}, x - \frac{1}{2}y)  (\Delta_FJ)^*(t- \frac{s}{2}, x - \frac{1}{2}y)\right]. \label{pxttt}
\eey
Here $\Delta_FJ(t, x)$ stands for  $\int dt' dx' \Delta_F(t - t', x- x') J(t', x')$, where
\bey
\Delta_F(t, x) = \int_{-\infty}^{\infty} \frac{dp}{4\pi \epsilon_p} e^{ipx - i \epsilon_p|t|},
\eey
is the Feynman propagator. The reason for the non-causal terms in the probability assignment is that the Feynman propagator does not vanish outside the light-cone.

The causal solutions to the  Klein-Gordon equation with a source, $\box \phi - m ^2 \phi = J$ are retarded solutions, i.e., they are generated  from $J$ through the retarded propagator, $\phi(t, x) = \int dx' \Delta_{ret}(t', x') J(t, x)$. The retarded propagator $\Delta_{ret}$ is simply the imaginary part of the Feynman propagator
\bey
\Delta_{ret}(t, x) = \int_{-\infty}^{\infty} \frac{dp}{4\pi \epsilon_p} e^{ipx}  \sin( \epsilon_p|t|),
\eey
and it vanishes outside the light-cone, i.e., for $|x| > |t|$.

In classical field theory, the restriction to retarded solutions is implemented as a boundary condition. This condition conflicts the time-reversal invariance of the evolution equation. Fundamentally, it is justified either cosmologically, or from thermodynamic time asymmetry ---see Chap. 2 of \cite{Zeh}.

Certainly, if we substitute $\Delta_F$ with $\Delta_{ret}$ in Eq. (\ref{pxttt}), the problem of causality is resolved\footnote{Analogous substitutions have been proposed in the context of photo-detection theory, in order to cure Glauber's model from the transient super-luminal terms \cite{phc1, phc2}.}. This amounts to a substitution of $e^{-i(\epsilon_p - \epsilon_{p'})t}$ with $\sin[(\epsilon_p - \epsilon_{p'})|t|]$ in Eq. (\ref{ptx4}). We cannot normalize by integrating $t$ along the whole real axis, because the error is exactly equal with that from the substitution of $\Delta_F$ with $\Delta_{ret}$. (They do not cancel.) )Still, for a large class of initial states, the correction in the total probability is small, so it is meaningful to define the post-selected density matrices  (\ref{posts}) and (\ref{rpp1}), and write the causally corrected version of Eq. (\ref{lpp}),
\bey
P_c(t, L) = \left| \int_0^{\infty} dp \psi_{ps}(p) \sqrt{v_p} e^{ipL - i \epsilon_pt}\right|^2  - \left| \int_0^{\infty} dp \psi_{ps}(p) \sqrt{v_p} e^{ipL + i \epsilon_pt}\right|^2, \label{pcx2}
\eey
modulo the fact that the probability density (\ref{pcx2}) is not normalized to unity.

However, the substitution of $\Delta_F$ with $\Delta_{ret}$, is equivalent to the substitution of the evolution operator  $e^{-i\hat{H}_{\phi}t}$ for the field with the operator $\hat{K}_t = e^{-i\hat{H}_{\phi}t} - e^{i\hat{H}_{\phi}t}$ which is non-unitary. Not only is such a substitution completely {\em ad hoc}, it also contradicts the basic rules of quantum theory. Furthermore, $\hat{K}_t$  makes no sense as the evolution of an initial state, because $\hat{K}_ 0 = 0$.
 As long as we use a measurement model in which the instant of detection is determined by the Hamiltonian, the sources are Feynman-propagated, and probabilities cannot be cured from super-luminal transients. A resolution requires at the very least, an incorporation of time-irreversibility in the description of measurement, in order to obtain some version of causal propagation.

\subsection{Causal propagation versus restricted propagation}

Next, we will describe the QTP approach to quantum field measurements \cite{AnSav12, AnSav19, AnSavHu22}, in which the space-time coordinates are genuine densities and the interaction between system and apparatus is always present. The key point here is that the propagator for an initial source is not the Feynman propagator. We will explain the benefits of this construction, and its potential for resolving the spurious superluminality problem from first principles. The resolution would require an  explicit modeling of the macroscopic apparatus in a way consistent with thermodynamic irreversibility. This is taken up in a different publication.

In QTP, we   describe measurement events as  transition between two complementary subspaces ${\cal H}_+$ and ${\cal H}_-$ of a Hilbert space  ${\cal H} = {\cal
H}_+ \oplus {\cal H}_-$. The subspace ${\cal H}_+$ describes the states of the system that are compatible with the realization of  the event under consideration is realized. If the event is a detection of a microscopic particle by  a measuring  apparatus, then  the subspace ${\cal H}_+$ corresponds to all states of the apparatus compatible with a macroscopic detection record.  We
denote  the projection operator onto ${\cal H}_+$ as $\hat{P}$ and the projector onto ${\cal H}_-$ as $\hat{Q} := 1  - \hat{P}$.

Transitions that are correlated with the emergence of a macroscopic record of observation are {\em logically
 irreversible}. Once they occur, and a measurement outcome has been recorded,   further time evolution of the system does not affect our knowledge.

After the transition has occurred,  a pointer variable $\lambda$ of the measurement apparatus takes a definite value. Let
$\hat{\Pi}(\lambda)$ be  positive operators that correspond to the different values of $\lambda$.  For example,
when considering transitions associated with particle detection, the projectors $\hat{\Pi}(\lambda)$  may be correlated  to the position, or to the momentum of the microscopic particle. Since $\lambda$ has a value only under the assumption that a detection event has occurred, the alternatives
  $\hat{\Pi}(\lambda)$ span the subspace
  ${\cal H}_+$ and not the full Hilbert space ${\cal H}$. Hence,   $\sum_\lambda \hat{\Pi}(\lambda) =
\hat{P}$.

Assuming that the systems is prepared in the state $|\Psi_0\rangle \in {\cal H}_-$, and that the Hamiltonian is $\hat{H}$, the probability distribution $P(\lambda, t)$ that the transition took place at time $t$ and an outcome $\lambda$ was recorded is
\begin{eqnarray}
P(\lambda, t) = \int d\tau  \langle \Psi_0|   \hat{C}^{\dagger}(\lambda, t- \frac{\tau}{2}) \hat{C}(\lambda, t+\frac{\tau}{2})
|\Psi_0\rangle, \label{pp2}
\eey
  where   the {\em class operator}
 \begin{eqnarray}
  \hat{C}(\lambda, t) := e^{i \hat{H}t} \sqrt{\hat{\Pi}}(\lambda)
\hat{H} \hat{S}_t \label{class}
\end{eqnarray}
is defined in terms of the restricted evolution operator $\hat{S}_t$ which is the continuous limit of the product $\hat{Q}e^{-i\hat{H} t/N} \hat{Q} e^{-i\hat{H}t/N} \hat{Q} \ldots \hat{Q}   e^{-i\hat{H}t/N} \hat{Q} $ as the number of steps $N$ goes to infinity. By the Mishra-Sudarshan theorem \cite{MiSu}, $\hat{S}_t$ is a unitary operator on ${\cal H}_-$ in this limit. However, one may modify the definition, for example, by regularizing the product so that there is an effective minimal time $\tau$. Then,  $\hat{S}_t$ may be non-unitary.

It is important to emphasize Eq. (\ref{pp2}) is a genuine probability density with respect to both $\lambda$ and $t$. The derivation of Eq. (\ref{pp2}) requires the assumption of decoherence in the measuring apparatus.

We note that $P(\lambda, t) $ vanishes if
 for $[\hat{P}, \hat{H}] = 0$. We consider a Hamiltonian   $\hat{H} = \hat{H_0} + \hat{H_I}$, where
$[\hat{H}_0, \hat{P}] = 0$, and $H_I$ a perturbing interaction. Since all transitions are due to $\hat{H}_I$,
to leading order in the perturbation,
\begin{eqnarray}
 \hat{C}(\lambda, t) = e^{i \hat{H}_0t} \sqrt{\hat{\Pi}}(\lambda) \hat{H}_I e^{-i \hat{H}_0t},
\label{perturbed}
\end{eqnarray}
When we use Eq. (\ref{perturbed}) for the measurement model of Sec. 2.1, and we take $\lambda$ to coincide with the position $x$ of a record,
we obtain Eq. (\ref{ptx4}). The derivation is conceptually more rigorous here, because Eq. (\ref{pp2}) is a genuine probability density to all orders of perturbation theory. Furthermore,  the interaction is always present, no switching is necessary in order to specify the instant of detection.

There are two important benefits in the QTP approach to measurements. We can employ use a dressed state for the apparatus, i.e., take a non factorized initial state $|\Psi_0\rangle$. In perturbation theory, $|\Psi_0\rangle = |\Omega, \psi_0\rangle + |\chi\rangle$, where $|\chi\rangle$ is a correction to leading order in the coupling. A first-principles identification of $|\chi \rangle$ will allow us to identify the vacuum noise for the measurement. Thus, we will be able to demonstrate whether the super-luminal transients are strongly dominated by noise, and hence, are unobservable.

It is highly unlikely that the non-factorized initial state will make the transients vanish. However, when taking an initial state
 Still, the lowest order contribution suffers from the super-luminal transients. However, fundamentally the initial state is propagated by a restricted propagator $\hat{S}_t$. For states derived by a localized source, this means that the source $J(x, t)$ is Feynman-propagated only at the tree level; higher order terms incorporate contributions from the apparatus through the projector $\hat{Q}$. We conjecture that these contributions will introduce  desired time-asymmetry in the evolution of $J(x, t)$, that will lead to the suppression of the transient terms.


 \section{Conclusions}

We presented our motivation and our results in the introduction. Here, we want to iterate the importance of  a consistent formulation of a QFT measurement theory, in order to provide a first-principles construction of relativistic quantum information \cite{AnSav22, AnSavHu22}. Such formulations are also important for the consistent description of proposed quantum optics  experiments in space \cite{Rideout, DSQL}. These are characterized by very long baselines, and they can explore fundamental issues of relativistic locality and causality at a fundamental level.

The issues analyzed in this paper are challenges to any  QFT measurement theory. Any such theory must provide specific predictions for time-of-arrival experiments, because in most of our measurements, the location of the detector is fixed and detection time is a random variable. It must also be consistent with locality and causality, meaning that it must provide an explicit recipe for constructing initial states, so that superluminal transients are either removed or they are drowned by vacuum noise. The first challenge is met by the QTP approach, the second is work in progress.

\section*{Acknowledgements}
CA acknowledges support by   grant  JSF-19-07-0001 from the Julian Schwinger Foundation; MEP acknowledges a studentship grant by the Moraitis School.

\end{document}